\documentclass[aps,prd,groupedaddress,nofootinbib]{revtex4}
\usepackage[english]{babel}
\usepackage{amsmath}
\usepackage{amsfonts}
\usepackage{amssymb}
\usepackage{latexsym}
\usepackage[mathcal]{eucal}

\usepackage{hhline}

\usepackage[colorlinks=false,breaklinks=true,plainpages=false]{hyperref}

\begin{document}

\title{Cosmological and Astrophysical Neutrino Mass Measurements\footnote{Prepared by attendees of the  workshop ``The Future of Neutrino Mass Measurements: Terrestrial, Astrophysical, and Cosmological Measurements
in the Next Decade,'' a program of the Institute for Nuclear Theory in Seattle in February, 2010.}}
\author{K.~N.~Abazajian$^1$, E.~Calabrese$^2$, A.~Cooray$^3$, F.~De Bernardis$^3$, S.~Dodelson$^{4,5,6}$, A.~Friedland$^7$, G.~M.~Fuller$^8$, S.~Hannestad$^9$, B.~G.~Keating$^8$, E.~V.~Linder$^{10,11}$, C.~Lunardini$^{12}$, A.~Melchiorri$^{2}$, R.~Miquel$^{13,14}$,E.~Pierpaoli$^{15}$, J.~Pritchard$^{16}$, P.~Serra$^{17}$, M.~Takada$^{18}$, Y.~Y.~Y.~Wong$^{19}$}

\affiliation{$^1$Maryland Center for Fundamental Physics, Department of Physics, University of Maryland, College Park, Maryland  20742  USA}
\affiliation{$^2$Physics Department and INFN, Universita' di Roma ``La Sapienza'', P.le Aldo Moro 2, 00185, Rome, Italy}
\affiliation{$^3$Department of Physics and Astronomy, University of
California, Irvine, CA 92697}
\affiliation{$^4$Center for Particle Astrophysics, Fermi National Accelerator Laboratory, Batavia, IL~~60510}
\affiliation{$^5$Department of Astronomy \& Astrophysics, The University of Chicago, Chicago, IL~~60637}
\affiliation{$^6$Kavli Institute for Cosmological Physics, Chicago, IL~~60637}
\affiliation{$^7$Theoretical Division, T-2, MS B285, Los Alamos National Laboratory, Los Alamos, New Mexico 87545, USA}
\affiliation{$^8$Center for Astrophysics and Space Sciences, Department of Physics, University of
California, San Diego, La Jolla, CA 92093-0424, USA}
\affiliation{$^9$Department of Physics and Astronomy,
 Aarhus University, DK-8000 Aarhus C, Denmark}
\affiliation{$^{10}$Berkeley Lab \& University of California, Berkeley, CA 94720}  
\affiliation{$^{11}$Institute for the Early Universe WCU, Ewha Womans University, 
Seoul, Korea} 
\affiliation{$^{12}$Arizona State University, Tempe, AZ 85287-1504, USA}
\affiliation{$^{13}$Instituci\'o Catalana de Recerca i Estudis Avan\c{c}ats, E-08010 Barcelona, Spain}  
\affiliation{$^{14}$Institut de F\'{\i}sica d'Altes Energies, E-08193 Bellaterra (Barcelona), Spain}
\affiliation{$^{15}$Department of Physics and Astronomy, University of Southern California, Los Angeles, CA 90089-0484, USA}
\affiliation{$^{16}$Harvard-Smithsonian Center for Astrophysics, MS-51, 60 Garden St, Cambridge, MA 02138}
\affiliation{$^{17}$Astrophysics Branch, NASA-Ames Research Center, 245-6, Moffett Field, CA 94035}
\affiliation{$^{18}$Institute for the Physics and Mathematics of the Universe  (IPMU), The University of Tokyo, Chiba 277-8582, Japan}
\affiliation{$^{19}$Institut f\"ur Theoretische Teilchenphysik und Kosmologie, RWTH Aachen, D-52056, Germany}


\date{\today}

%

\begin{abstract}
Cosmological and astrophysical measurements provide powerful constraints on neutrino masses complementary to those from accelerators and reactors. Here we provide a guide to these different probes, for each explaining its physical basis, underlying assumptions, current and future reach.
\end{abstract}

\maketitle

%

%


\section{Introduction}

Neutrinos are an integral part of the Standard Model of particle physics and are copiously produced by a variety of astrophysical sources.
Neutrinos also constitute a fraction of the dark matter in our Universe, leaving a characteristic imprint on various cosmological observables.
It is not surprising then that astrophysics and cosmology are poised to contribute to some of the most pressing problems in particle physics: understanding the properties of neutrinos, pinning down their masses, and ultimately understanding the origin of these tiny masses.

Recent advances in observational cosmology have resulted in tight constraints on the sum of the neutrino masses, 
and upcoming experiments are expected to improve on these results thanks both to new observational techniques and to larger data sets.
In this paper, we review the physical basis of the cosmological tests, the set of astrophysical and cosmological observables that are sensitive to neutrino properties and, for each, present current and future constraints. Lurking ahead is the tantalizing possibility of a detection: oscillation experiments have determined that the difference of the square of two of the neutrino masses is greater than $(0.05\, {\rm eV})^2$, implying a lower limit on the quantity that cosmological observations are most sensitive to: the sum of the neutrino masses. Current {\it upper bounds} range from a factor of 4-10 above the lower limit, so the grand challenge for the next generation of cosmological surveys -- detecting the effect of massive neutrinos on the cosmos -- appears within reach.


\section{Physical Basis of Cosmological Probes}

The Standard Models of particle physics and cosmology make a robust
prediction that the number density of relic neutrinos is $112$
cm$^{-3}$ per species.
This result is based solely on standard model physics and implies that massive neutrinos constitute a fraction
\begin{equation}
f_\nu = \frac{\Omega_\nu}{\Omega_m} = \frac{\sum m_\nu}{93 \, \Omega_m h^2 \, {\rm eV}} \simeq 0.08 \, \frac{\sum m_\nu}{1\,{\rm eV}}
\end{equation}
of the total matter density in the universe, where $\Omega_\nu$ and $\Omega_m$ are the neutrino 
energy density and matter density, respectively, in units of the critical 
density.  The last equality assumes the WMAP-7 best fit value, $\Omega_m h^2 = 0.134$ \cite{kom10}, where the Hubble rate today is parameterized as
$H_0\equiv 100 h$ km s$^{-1}$ Mpc$^{-1}$. The distribution of matter in the Universe depends sensitively on $f_\nu$, and therefore current and upcoming surveys that probe this structure in a variety of ways have the potential to constrain or measure the sum of the neutrino masses.

The small masses of neutrinos distinguish them from the rest of the matter in the Universe. Neutrino thermal velocities are non-negligible in the early universe and lead to smearing out of over-dense regions. At cosmic time $t$, neutrinos can free-stream distances of order $vt\sim (T_\nu/m_\nu)\times (1/H)$ where $H$ is the Hubble rate and $T_\nu$ the neutrino temperature, calibrated from the Cosmic Microwave Background (CMB) to be $1.9a^{-1}$K where $a$ is the scale factor (set equal to one today) governing the expansion of the Universe. The {\it comoving} free-streaming length scale is therefore $vt/a\sim 0.04 f_\nu^{-1} h^{-1} a^{-1/2}$ Mpc. Neutrinos do not clump on scales significantly smaller than this free-streaming scale. When any component of the density does not clump, the delicate balance between dilution due to the expansion of the Universe and accretion due to gravitational instability is upset, and gravitational potential wells decay. Over the course of billions of years, this decay is appreciable even if only a small fraction of the matter is not participating in the cosmic dance of structure formation. Structure on scales smaller than $\sim0.1$ Mpc/$f_\nu$ is suppressed for all $a$, while
scales larger than ~$100$ Mpc are never affected. Neutrinos therefore produce a characteristic fall-off in the {\it power spectrum} of the matter distribution from large to small scales. In linear perturbation theory the suppression of power is roughly given by $\Delta P/P \sim -8 f_\nu$, and if non-linear corrections are included the suppression increases to $\Delta P/P \sim -10 f_\nu$ for Fourier modes with wavenumber $k \sim 0.5-1 \, h \, {\rm Mpc}^{-1}$
\cite{Brandbyge:2008rv,Brandbyge:2009ce,Viel:2010bn,Wong:2008ws,Lesgourgues:2009am,Saito:2009ah}.

There are a wide variety of cosmological probes of the matter distribution, each of which has the potential to detect the signature suppression caused by neutrino masses. Dozens of surveys over the coming decade will make detailed observations, hidden in which will be clues to the neutrino mass. Extracting the relevant information will be challenging: a combination of insight and improved computational capability will be necessary to confront simply the theoretical systematics that threaten to obscure the signal.  However, the possibility of detecting the signature of neutrino masses in cosmology is so alluring that scores of researchers are devoted to address the most pressing issues. The wide variety of probes is absolutely essential since each has its own set of strengths and weaknesses.
Further such a joint analysis  of different probes covering wider ranges of
redshifts and distance scales measured will be a powerful way of
efficiently breaking parameter degeneracies. For example, both dark energy and neutrino mass 
suppress structure formation, but leave their imprints on different sets of length scales and redshifts.

\section{Probes}

There are two main techniques for probing the matter distribution: mapping the distribution of {\it biased}
tracers and observing the subtle effects of gravitational lensing. The most traditional tracer is the
galaxy distribution, which is related to the underlying matter distribution by a bias factor that can be both time
and scale dependent (although it is likely constant on large scales). Other biased tracers are neutral hydrogen
-- as mapped by Lyman $\alpha$ absorption or by 21 cm emission -- and galaxy clusters. Gravitational
lensing is different in that it is sensitive to the gravitational potential directly (which is linearly related
to the matter distribution by the Poisson equation), but lensing of objects at a given distance from us depends
on all values of the potential along the line of sight so offers only a 2D, projected view of the distribution.
Table \ref{tab1} lists these probes, the limits currently obtained and those that might be reached with future
surveys.%
\footnote{Note that the potential of a future laboratory tritium $\beta$-decay experiment to probe
the absolute neutrino mass scale is usually characterized by two numbers: its {\it sensitivity} to the neutrino
mass, defined as the 95\% upper limit the experiment can set on the neutrino mass if the true neutrino
mass is zero, and its {\it discovery potential} (or detection threshold), defined as the minimal mass that the
neutrino should have in order for the mass to be detected by the experiment at some confidence level (say,
95\%). The future cosmological limits presented in Table~\ref{tab1} have been derived formally as
95\% sensitivities to $\sum m_\nu$.  However, for reasonable cosmological models, the sensitivity and the
95\% discovery potential for a given probe are generally numerically
quite similar~\cite{Hannestad:2007cp}.
Therefore, as a rule of thumb,
the numbers denoted ``reach'' in Table~\ref{tab1} can be taken to mean both the
sensitivity and the discovery potential at 95\% C.L.}

Each of these probes faces technological, observational, and theoretical challenges in its quest to extract a
few percent level signal. Table~\ref{tab1} highlights the key theoretical systematics each probe will have to
overcome to obtain a reliable constraint on neutrino masses.

\begin{table}[t]
\begin{center}
\begin{tabular}{|p{3cm}|p{1.7cm}|p{1.8cm}|p{3.cm}|p{3.cm}|p{3.5cm}|p{2.5cm}|}
\hline
Probe & Current $\sum m_\nu$ (eV) & Forecast $\sum m_\nu$ (eV)  & Key Systematics & Current Surveys &  Future Surveys\\
\hline
\hline
CMB Primordial & $1.3$&$0.6$ & Recombination & WMAP, Planck & None \\
\hline
CMB Primordial $+$ Distance & $0.58$&$0.35$ & Distance measurements & WMAP, Planck & None \\
\hline
Lensing of CMB & $\infty$&$0.2-0.05$ & NG of Secondary anisotropies &
Planck, ACT \cite{act}, SPT \cite{spt} & EBEX~\cite{ReichbornKjennerud:2010ja}, ACTPol, SPTPol, POLARBEAR~\cite{polar}, CMBPol
\cite{Baumann:2008aq} \\
\hline
Galaxy Distribution & 0.6&$0.1$ & Nonlinearities, Bias & SDSS
\cite{Reid:2009xm,Percival:2009xn},  BOSS \cite{boss} & DES \cite{des}, BigBOSS~\cite{bigboss}, DESpec~\cite{despec},
LSST \cite{:2009pq}, Subaru PFS~\cite{subaru}, HETDEX \cite{bib:hetdex} \\
\hline
Lensing of Galaxies & 0.6&$0.07$ & Baryons, NL, Photometric redshifts & CFHT-LS
\cite{Fu:2007qq}, COSMOS~\cite{cosmos} & DES \cite{des}, Hyper SuprimeCam, LSST~\cite{:2009pq}, Euclid
\cite{euclid}, WFIRST\cite{wfirst} \\
\hline
Lyman $\alpha$ & $0.2$&$0.1$ & Bias, Metals, QSO continuum & SDSS, BOSS,
Keck & BigBOSS\cite{bigboss}, TMT\cite{tmt}, GMT\cite{gmt}  \\
\hline
21 cm & $\infty$&$0.1-0.006$ & Foregrounds, Astrophysical modeling & GBT~\cite{gbt}, LOFAR \cite{lofar}, PAPER~\cite{Parsons:2009in}, GMRT~\cite{gmrt} &
MWA \cite{mwa}, SKA \cite{ska}, FFTT \cite{mao2008} \\
\hline
Galaxy Clusters & $0.3$&$0.1$ & Mass Function, Mass Calibration & SDSS, SPT, ACT, XMM~\cite{xmm} Chandra~\cite{chandra} &  DES, eRosita~\cite{rosita}, LSST\\
\hline
Core-Collapse Supernovae & $\infty$ & $\theta_{13}>0.001^*$ & Emergent $\nu$ spectra & SuperK~\cite{superk}, ICECube\cite{icecube} & Noble Liquids, Gadzooks~\cite{gadzooks} \\
\hline
\end{tabular}
\caption{\label{tab1}Cosmological probes of neutrino mass.  ``Current'' denotes published (although in some cases controversial, hence the range)  
95\% C.L/~upper bound on
$\sum m_\nu$ obtained from currently operating surveys, while ``Reach'' indicates the forecasted 95\%
sensitivity on $\sum m_\nu$ from future observations.  These numbers have been derived for a minimal
7-parameter vanilla+$m_\nu$ model. The six other parameters are: the amplitude of fluctuations, the slope of the spectral index of the primordial fluctuations, the baryon density, the matter density, the epoch of reionization, and the Hubble constant.\\
$^*$ If the neutrinos have the normal mass hierarchy, supernovae spectra are sensitive to $\theta_{13}\sim 10^{-3}$. The inverted hierarchy produces a different signature, but one that is insensitive to $\theta_{13}$.} 
\end{center}
\vspace{-0.6cm}
\end{table}

\subsection{Primordial Cosmic Microwave Background}

In the first row of Table~\ref{tab1}, we report the constraints obtained using 2-point statistics of the CMB:
temperature and polarization auto-spectra and the temperature-polarization cross-spectrum. Massive neutrinos increase the anisotropy
on small scales because the decaying gravitational potentials enhance the photon energy density fluctuation (see, e.g., 
\cite{mb,1996ApJ...467...10D}).
Also, the sound horizon, which dictates the position of the acoustic peaks, shifts due to the slightly different expansion history 
caused by massive neutrinos.
The current WMAP
7-year dataset constrains the
sum of neutrino masses to $1.3$ eV at $95 \%$ c.l.~\cite{kom10}
within the standard cosmological model, $\Lambda$CDM. Planck data alone will constrain $\Sigma
m_{\nu}$ to $0.6$ eV at $95 \%$ C.L. (see, e.g., \cite{deb09}).
This constraint should be considered as the most conservative and
reliable cosmological constraint on neutrino masses.
A tighter constraint on the neutrino masses can be obtained by
combining CMB observations with measurements of the Hubble constant
$H_0$ and cosmic distances such as from Type Ia supernovae and 
Baryon Acoustic Oscillations (BAO). The WMAP7+BAO+$H_0$ analysis of~\cite{kom10}
reports a constraint of  $0.58$ eV at $95 \%$ C.L., while a constraint
about a factor
$2$ smaller could be achieved when the Planck data will be combined
with similar datasets.

The key theoretical systematics in confronting the CMB predictions with data have been overcome. The physics is linear, so all codes agree with the requisite precision. Precise constraints require careful treatment of many of the excited states of hydrogen during recombination~\cite{Seager:1999bc}, but here too recent advances~\cite{AliHaimoud:2010dx} have attained the precision needed to extract accurate information from Planck.
There are uncertainties associated with the distance measurements given by $H_0$ and BAO, but again these seem to be under tighter control.

\subsection{Lensing of the CMB}

The cosmic microwave background radiation is gravitationally lensed
by matter inhomogeneities along the line of sight to the last scattering surface at $z_{\rm lss}=1090$.  Lensing affects the temperature and
the polarization of the CMB in several ways. First, the power spectra are smoothed out, an effect
that makes sense intuitively since random deflections tend to reduce the amplitude of hot/cold spots.
A fit for the presence of lensing
on the power spectrum gives a non-zero result (consistent with prediction)
at 3.4$\sigma$ \cite{Shirokoff:2010cs} and 2.8$\sigma$ \cite{acbar}. More dramatically, polarization maps can be decomposed into $E$- and $B$-modes, the latter of which is not
produced by (scalar) density perturbations. Lensing though transforms $E$-modes into $B$-modes with a characteristic spectral
shape that depends on the
integrated gravitational potential. This shape depends on the sum of the neutrino masses. 

The most powerful way to map the projected gravitational potential
is to measure CMB polarization on small scales. Each CMB photon is deflected by only a small
amount (of order a few arcminutes) but the structures responsible for lensing are coherent over degree scales.
This leads to a counter-intuitive probe: CMB structure on small scales offers information about the gravitational
potential on large scales. Extracting this information has been the subject of some elegant theoretical work~\cite{Hu:2001tn,Hirata:2003ka}
focused on the higher order moments of the temperature field.

Claims of detection of lensing in the higher-point functions
initially relied on cross-correlating with matter tracers (since the auto-correlation
that will eventually be so powerful is much noisier), with detections~\cite{smith,hirata} at the 3$\sigma$
level. Evidence for the auto-correlation in WMAP data~\cite{Smidt:2010by} was reported in 2010, followed by a $4\sigma$ detection in ACT~\cite{2011arXiv1103.2124D},
but no direct constraints on $\sum m_\nu$ from CMB
lensing exist to date.
Near-term (next three years) results should enable determination of
$\sum m_\nu$ to 0.2 eV (95\% cl) from the Planck satellite and ground-based
polarization experiments.  Long term results (15 years) from CMB lensing
(CMBPol/EPIC satellite) strive for 0.04 eV \cite{Baumann:2008aq}.

CMB lensing is similar to the galaxy lensing method, but has some advantages
and disadvantages relative to it \cite{Cooray:2002py}.  One key property is that the source
redshift ($z_{\rm lss}$) is accurately known, in contrast to galaxy
lensing where considerable effort is needed to characterize the sources.
Furthermore, the source (CMB) redshift is high, so CMB lensing probes the
matter (or potential) distribution at higher redshifts, $z\approx1$-4, exploring the
universe at an epoch different from many other cosmological techniques. Since large scales and high 
redshifts are probed, the density field is very nearly linear and non-linear
complications are not important.
%
On the other hand, the CMB source is at a single redshift, giving only a
single weighted measurement of the gravitational potential and distance
factors, rather than the redshift tomography possible with galaxy lensing. Perhaps 
the most serious systematic is the
impact of non-Gaussianities of other secondary anisotropies on the estimators used to extract the potential.

\subsection{Galaxy Distribution}

Galaxy surveys have until now been the most direct way of measuring the matter power spectrum on intermediate and small scales, and therefore also the most direct probe of the suppression of fluctuation power caused by the presence of massive neutrinos. At present by far the largest spectroscopic survey is the Sloan Digital Sky Survey (SDSS), and, together with the WMAP CMB data, it provides an upper bound of approximately 0.6 eV on $\sum m_\nu$ \cite{Reid:2009xm}.

Galaxy redshift surveys measure the power spectrum of
 galaxy number density fluctuations, $P_g(k)$.  In turn, this power
 spectrum is related to the underlying  matter power spectrum $P(k)$
via
\begin{equation}
P_g(k)=b^2(k) P(k),
\end{equation}
where the bias parameter $b$ depends on both scale and on the type of galaxies surveyed. This has been shown to be a significant problem for some surveys and therefore emphasis has shifted towards basing surveys on luminous red cluster galaxies which constitute a fairly homogeneous sample. This is for example the case for the SDSS-LRG sample which was used to derive the current 0.6 eV upper bound.

A number of larger galaxy surveys will be carried out within the next decade and will increase sensitivity to neutrino mass significantly.
Given some survey design one can expect to measure the galaxy power spectrum
$P_g(k)$
up to a statistical uncertainty of
\cite{Tegmark:1997rp}
\begin{equation}
\label{eq:error}
\Delta P_g(k) = \sqrt{\frac{1}{2 \pi  \ w(k) \ \Delta \ln k}} \left[P_g(k) + \frac{1}{\overline{n}_g} \right].
\end{equation}
Here, $w(k) = (k/2 \pi)^3 \ V_{\rm eff}$, $V_{\rm eff}$
is the effective volume of the survey, $\overline{n}_g$ is the
galaxy number density, and $\Delta \ln k$ is the bin size at $k$ in $\ln k$-space. Future surveys will go deeper (and therefore survey even fainter galaxies, leading to larger $\overline{n}_g$) and wider (and therefore increased $V_{\rm eff}$) leading to much smaller errors on the power spectrum.

The precision with which the power spectrum can, in principle, be measured is related to the survey volume because that is a measure of the number of independent Fourier modes available. On small scales precision is limited by shot noise, i.e.\ by the sparseness of galaxies.
However, in practice this is not the most significant problem on small scales. Rather, the usefulness of small scale data is limited by the fact that structures are non-linear. At $z = 0$ this effectively cuts away all data at $k > 0.1 \, h$/Mpc. 
Additionally, the luminosity dependence
of the galaxy bias and its evolution can be combined with clustering information and the CMB primary spectrum to constrain the neutrino masses \cite{Seljak:2004sj}. The current constraint on the neutrino masses with this method using a range of galaxy clustering data from SDSS, DEEP-2 at $z \sim$ 1, and Lyman-break clustering at $z \sim3$ is 0.28 eV at 95\% C.L \cite{DeBernardis:2008qq}.

However, most upcoming surveys aim at measuring at higher redshift than the SDSS and therefore the problem of non-linearity will be somewhat alleviated. In \cite{Hannestad:2007cp} a study of neutrino mass constraints was carried out for a number of proposed surveys combined with Planck data.
Very roughly, the HETDEX \cite{bib:hetdex} or BOSS \cite{boss} surveys, together with Planck should push the sensitivity to about 0.2 eV at 95\% C.L., and a future space-based mission such as WFIRST or EUCLID could yield a sensitivity of around 0.1 eV (95\% C.L.).

The major theoretical hurdles that need to be addressed in order to extract these sensitive limits are understanding the nonlinearities and bias. Simulations and cross-correlating with lensing surveys can help with these issues.

\subsection{Lensing of Galaxies}

Weak gravitational lensing (or cosmic shear) of distant galaxies by the intervening
large scale structure provides an elegant way to map directly the matter
distribution in the universe.
Perturbations in the matter density field
between the source and the observer bend the paths of light rays, thereby inducing
distortions in the observed images of source galaxies.
By measuring the angular correlation of these distortions, one can
probe the clustering statistics of the intervening matter density field.
This again allows a probe of the neutrino masses \cite{Cooray:1999rv}.

Current weak lensing surveys are already providing interesting constraints on neutrino masses.
An analysis of the CFHTLS data from a 30 square degree sky patch
finds a 95\% C.L. upper limit of $\sum m_\nu < 1.1$~eV in
a 7-parameter vanilla+$m_\nu$ model
when combined with the WMAP 5-year data~\cite{Ichiki:2008ye}.  A tighter
constraint, $\sum m_\nu < 0.54$~eV, is obtained when distance measurements from SNIa and BAO
are also included
in the analysis~\cite{Ichiki:2008ye}.

Future dedicated lensing surveys will probe higher redshifts with
almost
full sky coverage.
Furthermore, all surveys will provide photometric redshift information on the source galaxies.
This additional information allows for the binning of galaxy images by redshift and hence tomographic
studies of the evolution of the intervening large scale structure and the distance-redshift relation.
The LSST~\cite{:2009pq} combined with primary CMB anisotropy measurements from Planck
can constrain $\sum m_\nu$ down to $\sim 0.07$~eV (95\% C.L.) using five tomography bins~\cite{Hannestad:2006as}.
Similar sensitivity is expected also for Euclid~\cite{Kitching:2008dp}.

{\it Dominant systematics}.  On the observational side,
photometric redshift measurements typically have uncertainties of $\Delta z=0.03 \to 0.1$.  Accurate
modeling of this uncertainty will be important for tomographic studies. The measurements themselves of course require
great care and much work has been done over the last decade understanding how to use the stars to correct for instrumental and atmospheric distortions.
On the theory side, future weak lensing surveys will derive most of their constraining power from nominally nonlinear scales
$k > 0.1 \ {\rm Mpc}^{-1}$.  This will require that we control the uncertainties in our theoretical predictions of the
nonlinear power spectrum to
a
 percent level.  Baryon physics will also be important here.  The study
of~\cite{Jing:2005gm} finds that baryon physics can contribute an uncertainty of up to 10\% at multipole $\ell > 1000$ corresponding to physical scales $k$ less than a factor of ten beyond the linear regime.

\subsection{Lyman $\alpha$ Forest}

The expectation that cosmological intergalactic low-density gas
follows the gravitationally dominant dark matter near the nonlinear
clustering scale has led to the proposal of measuring structure in
dark matter clustering via the absorption features along the line of
sight to a distant quasar, namely, through the Lyman-$\alpha$ forest (e.g.,~\cite{Croft:2000hs}).  
The results produced some of the most
sensitive results on the amplitude and shape of dark matter clustering
at small scales, and therefore indirectly on the presence of massive
neutrinos when combined with the CMB, though that was not derived in
the initial results. This method was immediately questioned due to the
effects of a smoothing introduced by peculiar velocities in the
forest, as well as uncertainties in the ionizing background of the
gas~\cite{Gnedin:2001wg}.  The flux power spectrum was also shown to
be affected by fluctuations in the temperature of the intergalactic
medium (IGM)~\cite{Lai:2005ha}, the temperature and ionization history
of the gas, and metal line contamination~\cite{Jena:2004fc}.
The promise lies in the sensitivity of the Lyman-$\alpha$ forest to dark matter clustering on small scales, 
where massive neutrinos would suppress the power spectrum.
The measurement of small-scale power shape and amplitude provides information on
neutrinos only in combination with precise measures of the large scale
matter clustering amplitude from the cosmic microwave background
(CMB), from WMAP or Planck (for a review, see, e.g.,
Ref.~\cite{Lesgourgues:2006nd}).

An important systematic is the {\it bias} relating the power spectrum of the flux (the observable)
to the power spectrum of the matter.
This bias is sensitive
to the temperature history and assumed temperature-density relation of the gas 
~\cite{Peeples:2009uj,Peeples:2009ue,Abazajiandegen}.  So far,
inversions from the distribution of gas to dark matter have used rigid
power-law evolution prescriptions for the temperature-density relation
of the gas and its evolution over cosmic history from redshifts $z=1 -
7$.  The parameterized power-law functional forms for the evolution of
the temperature density relation were constrained simultaneously with
the inferred matter power spectrum, within the inversion of the
gas-to-matter bias in the flux power spectrum.  Such work also
typically includes priors from independent measures of the temperature
density relation at specific redshifts, leading to very tight
constraints on the global temperature density relation evolution as
well as other parameters~\cite{McDonald:2004xn,Viel:2005ha}.

Setting potential shortcomings of the method of inversion of the
gas-matter bias relation aside, the constraints arising from the flux
power spectrum from Sloan Digital Sky Survey
quasars~\cite{McDonald:2004xn,Viel:2005ha} are quite stringent.  They
are stringent due to the quoted small intrinsic errors on the shape
and amplitude of the inferred matter power spectrum, but also because
this initial work found that the amplitude of the inferred matter
power spectrum was in tension with the
WMAP 3-year
results~\cite{Seljak:2006bg}, leading to a small and more stringent
region where likelihoods for the amplitude and shape of the dark
matter power spectrum were consistent (Ref.~\cite{Seljak:2006bg},
Fig.~1).  The resulting constraint on the neutrino mass is the most aggressive
to date: $\sum
m_{\nu}\le 0.17\rm\ eV$ (95\% CL).  An independent analysis of the Lyman-$\alpha$
forest flux power spectrum is finding a residual correlation between
spectral noise and the inferred primordial flux power, that may lead
to a large change in the inferred matter power spectrum shape and
amplitude~\cite{Abazajianflux}.
Finally we also note the recent and very detailed study performed by 
\cite{Viel:2010bn} in which neutrinos were directly included in the N-body simulations from which 
the flux power spectra were calculated. They derive a bound based on WMAP-7 plus SDSS Lyman-$\alpha$ of 0.9 eV (95\% C.L.).

Future forecasts of the combination of high-resolution Keck and/or VLT
spectra of the Lyman-alpha forest show that the sensitivity level of
Lyman-$\alpha$ forest measures can be quite stringent when combined
with the CMB from Planck, reaching potential sensitivities of $\sum
m_{\nu}\le 0.11\rm\ eV$ (95\% CL)~\cite{Gratton:2007tb}.  These forecasts
employ simplified models for the temperature
history of the gas and global temperature-density relation evolution
over cosmic time, therefore it is not certain how a more general
approach would change forecast sensitivities.

\subsection{21 cm Surveys}

Low frequency radio observations of the redshifted 21 cm line of neutral hydrogen map have the potential to map the distribution of matter at high redshifts measuring the matter power spectrum just as galaxy surveys do at low redshift.  Two main epochs can be probed with different designs of array: the epoch of reionization (EoR) at $6\lesssim z\lesssim 12$, where neutral hydrogen is present in the intergalactic medium, and intensity mapping at $z\lesssim4$ where the neutral hydrogen in dense clumps is targeted. The signal to be observed is 4-5 orders of magnitude smaller than foreground synchrotron emission from the Galaxy, making foreground removal the biggest issue for the success of these surveys.  Fortunately, the smoothness of the foregrounds can be exploited to aid removal and there is cause to be optimistic.

21 cm EoR experiments measure fluctuations in the 21 cm brightness temperature $T_b$ that can be sourced by fluctuations in the ionization field ($x$), arising from the ionized hydrogen bubbles around clusters of galaxies \cite{trac2009}, as well as fluctuations in the density so that the observed power spectrum includes both contributions \cite{fob}
\begin{equation}
P_{21}(k,\mu)=T_b^2(P_{\delta\delta}+2P_{x\delta}+P_{xx}+2\mu^2(P_{\delta\delta}-P_{x\delta})+\mu^4P_{\delta\delta}),
\end{equation}
where redshift-space distortions induce an angular dependence, 
with $\mu=\hat{\mathbf k}\cdot \hat{\mathbf n}$ the angle between a Fourier 
mode and the line of sight. If the contribution from the ionized regions can be accurately modeled, cosmological constraints are possible.  Otherwise, more precise measurements to use the angular dependence to separate the density contribution are required.

Since EoR experiments measure the power spectrum at high redshifts they can probe very large volumes where the non-linear scale is small, making many Fourier modes accessible for very precise parameter constraints.  At $z=8$ the non-linear scale is $k_{\rm nl}=2{\,\rm Mpc^{-1}}$, almost an order of magnitude larger than at $z=0$.  The sensitivity of the instruments is determined primarily by their collecting area $A_{\rm tot}$ and by the number of antennae $N_{\rm ant}$ correlated together by the interferometer.  Initial pathfinder instruments (MWA, LOFAR, PAPER, GMRT) have yet to claim a detection. Even if they do succeed, precision constraints will require the proposed Square Kilometer Array (SKA).  Clever experimental design \cite{tegmark2008} may allow use of the FFT to efficiently correlate many dipoles, increasing $N_{\rm ant}$ dramatically and leading to an even more sensitive instrument dubbed the Fast Fourier Transform Telescope (FFTT).

Predictions for neutrino mass constraints suggest that, in combination with Planck, MWA will constrain $\sum{m_\nu}\approx0.1{\, eV}$, SKA will constrain $\sum{m_\nu}\approx0.02{\, eV}$, while the FFTT could reach $\sum{m_\nu}\approx0.003{\, eV}$ \cite{mcquinn2006,mao2008}.  The sensitivity of FFTT is sufficient that, in principle, individual neutrino masses could be measured at low significance \cite{pritchard2008}.  The constraints quoted above degrade by a factor of $\sim2$ when reionization modeling is required and only FFTT is capable of useful constraints ($\sum{m_\nu}\approx0.02{\, eV}$) if the angular decomposition is required \cite{mao2008}.

21 cm intensity mapping complements optical galaxy surveys at $z\lesssim3$.  Rather than identifying individual galaxies and then binning to estimate the density field, these experiments integrate the entire 21 cm flux emitted from the neutral hydrogen in galaxies within the beam.  This sacrifices high angular resolution, most of which probes non-linear scales, for increased survey speed and reduced cost.  In addition to being affected by the usual issues of galaxy bias and non-linearity on small scales, these experiments must worry about the possibility of large scale variations in the ionizing background, which could modulate the neutral hydrogen power spectrum (although recent estimates suggest this is a small effect \cite{wyithe2009}).  Possible neutrino mass constraints from post-reionization 21 cm experiments are considered in \cite{visbal2009}.  For their proposed experiments MWA-5k and FFTT targeted at $z=3.5$ in combination with Planck they find $\sum{m_\nu}
 =0.02{\, eV}$ and $0.04{\, eV}$ respectively if constant bias is assumed.  An initial application of intensity mapping at $z=0.8$ has recently been carried out \cite{gbt}.  

\subsection{Galaxy Clusters}

Galaxy clusters, with masses around $10^{14}-10^{15} M_{\rm solar}$ are the largest gravitationally bound objects in our Universe.
Observations of cluster number counts in a given volume of the Universe provide information on
the amplitude of inhomogeneities on a range of scales around $k \simeq 0.1 h Mpc^{-1}$ and therefore are sensitive to neutrino mass.

Clusters  are observed in different wavebands: in the radio and X--ray due to intra-cluster gas emission and in the optical/infrared from their galaxies.
Optical and X--ray observation have been performed and exploited for cosmological purposes in the past decades using a few hundreds of clusters, while  major surveys in the radio band are starting to produce results at this time.  The radio surveys exploit the Sunyaev-Zel'dovich (SZ) effect from inverse Compton scattering, which has the advantage that the radio signal of clusters does not become dimmer with redshift.  Future optical surveys like LSST are expected to detect a similar number of clusters.

A key issue which arises when confronting observations with theory is: while the latter makes quite precise predictions for the
number of objects in a given mass range expected for a given cosmology,
the mass of a cluster is not directly observable and is related in a non-trivial way to 
the measured quantities (optical, lensing, X-ray, and SZ signals). 
This mass calibration uncertainty typically constitutes the largest source of systematic error for all types of cluster observations. The
standard approaches to address this issue rely on mass--observable relations calibrated either through
numerical simulations or through observations (exploiting either the hypothesis of hydrostatic equilibrium or lensing effects to derive  the mass).
While comparing results from different methods has overcome some of the individual
biases,  assumptions in the scaling relations are still the outstanding issue.
Future surveys delivering many more clusters at different redshifts will greatly help in addressing issues related to galaxy formation and evolution in clusters and understanding the physical processes  of the intra-cluster medium.

Current limits on neutrino mass from cluster number counts essentially come from X--ray observations.
Combining X--ray cluster number counts with WMAP, BAO, and SN data yields $\Sigma {m_\nu} \le  0.33 $ eV  at $95\%$ C.L. when also dark energy is considered, which is half the limit achievable without clusters \cite{Vikhlinin09}.
While attempts have been made to derive cosmological constraints form clusters observed in galaxy surveys like SDSS, limits on neutrino masses have not been derived yet.

Upcoming  surveys in the optical and in the radio yielding tens of thousands of clusters will also allow for the determination of the galaxy cluster power spectrum.
By combining information from Planck, the number counts and the power spectrum probes from radio surveys like SPT and optical ones like LSST may achieve a precision $\sigma ( \Sigma {m_{\nu}}) = 0.04-0.07$ eV \cite{Wang06}.

\subsection{Supernovae}

A core collapse supernova neutrino burst signal promises unique insights into the neutrino mass hierarchy and the mixing angle $\theta_{1 3}$. While the processes of gravitational potential growth and decay in the universe are sensitive to the cosmological neutrino background and, in particular, the neutrino mass, likely they are quite insensitive to neutrino flavor mixing. In contrast, the physics of massive star collapse and the neutrino signature from such an event is insensitive to absolute neutrino rest masses, but can be very sensitive to the neutrino mass hierarchy and flavor mixing.

Large scale numerical simulations show that there can be features in the supernova neutrino fluxes and energy spectra that have their origin in the nonlinear coupling of the flavor histories of outgoing neutrinos. Chief among these features is the spectral swap/split. In the normal mass hierarchy (NH), this swap feature manifests itself as nearly complete neutrino flavor transformation below a characteristic swap energy ($\sim 10\,{\rm MeV}$). In the inverted mass hierarchy (IH) the swap has an opposite sense, with most neutrinos above the swap energy transforming, while those with energies below the swap energy do not. This is a fairly dramatic signature that, if detected, would pin down the nature of the hierarchy. Moreover, in the NH the swap energy depends on $\theta_{1 3}$. This energy decreases as $\theta_{1 3}$ is decreased, suggesting that a detected supernova neutrino burst could give a measure of this unknown vacuum mixing angle. In the NH we would likely need $\theta
 _{ 1 3} >{10}^{-2} - {10}^{-3}$ to see the swap. In contrast, in the IH the swap phenomenon and the swap energy are very insensitive to $\theta_{1 3}$, with the full swap evident even for values of this mixing angle many orders of magnitude below what   could be probed in reactor and long baseline experiments.

The principal difficulties in extracting neutrino properties from a supernova neutrino burst revolve around limitations in numerical modeling and the experimental issues associated with the detection itself. Although the swap phenomenon itself probably is relatively robust, recent work makes clear that the nonlinear neutrino flavor transformation regime should be treated with a full $3\times3$ matrix and ``multi-angle'' computation. There are as yet only a few such calculations. These numerical studies are in a nascent stage, and there may be surprises as this work progresses. Core collapse supernovae in the Galaxy occur relatively infrequently (e.g., every 30 years or so), so an obvious question is whether the right kinds of detectors (e.g., water Cerenkov, liquid scintillator, etc.) will be around to catch a burst. Additionally, once the objective of detection becomes resolving a swap/split, the experimental scheme must focus on the relatively low expected swap energies. Liquid noble gas detectors being
crafted for dark matter detection might be configured to do this.

\section{Model depdendence}

Upper bounds on the neutrino mass --- and, indeed, any positive measurement thereof in the future --- from 
precision cosmological observations are inherently model-dependent. This dependence arises because
the presence of neutrino hot dark matter manifests itself as a relatively smooth feature in the cosmological observables;
Any information about the neutrino mass can be obtained only by way of statistical inference
from the observational data after a parametric model has been chosen as the basis for the analysis.

As of now there is a general consensus in the cosmology community that the
simplest model required to account for all cosmological 
observations is the concordance flat $\Lambda$CDM model. 
This model assumes 
\begin{enumerate}
\item General relativity holds on all length scales; 
\item The large-scale spatial geometry of the universe is flat; 
\item The Universe consists of (a) photons, whose 
 energy density is fixed by the COBE FIRAS measurement of the CMB temperature and energy spectrum~\cite{Fixsen:1996nj},
 (b)
three families of thermalized neutrinos whose temperature is 
linked to the CMB temperature via $T_\nu = (4/11)^{1/3} \, T_\gamma$ 
from entropy conservation arguments, and (c) 
atoms\footnote{The atomic part of the universe's energy budget is often labeled ``baryons'' by cosmologists, since
the non-baryonic part of an atom, namely the electrons, account only for a negligible fraction of the 
atom's total mass.},
cold dark matter\footnote{A dark matter fluid is classified as cold if it is non-relativistic and has negligible velocity 
dispersion in the epochs relevant for the generation of CMB anisotropies and structure formation.},
and vacuum energy due to a cosmological constant;

\item The initial conditions, i.e., the statistics and amplitude of the primordial perturbations to the 
Friedmann-Lema\^{\i}tre-Robertson-Walker metric, 
are set by the simplest single-field inflation models.  The perturbations are adiabatic, and minimal primordial 
gravitational waves are produced during inflation.
The spectrum of density perturbations is described by a single spectral index $n_s$ and an amplitude $A_s$, both of which are free 
to vary.

\end{enumerate}

Besides the parameters mentioned above, a minimal model of the CMB anisotropies requires an additional astrophysical parameter, the optical depth to reionization $\tau$.  
Furthermore, a number of nuisance parameters are used in the analysis of galaxy clustering data 
to describe certain nonlinear effects that are, as of now,  not fully understood or 
calculable from first principles.  These are marginalized at the end of the analysis.  
In all there are six free physical parameters in the so-called ``vanilla'' model of cosmology:
\begin{equation}
\Omega_b h^2, \Omega_{\rm CDM} h^2, \Omega_\Lambda, n_s, A_s,\tau.
\end{equation}
Spatial flatness implies $\Omega_b + \Omega_{\rm CDM} + \Omega_\Lambda=1$, from which 
we can deduce the reduced Hubble parameter $h$.  
The simplest neutrino mass limits can be obtained by including in this empirical description a variable amount of hot dark matter characterized by 
the sum of the neutrino masses $\sum m_\nu$.

A number of variations around this general framework are possible and generally fall into the following categories:
\begin{enumerate}
\item {\it Extra relativistic species:} \quad  Besides three light neutrinos, several particle physics puzzles have opened
up the possibility of non-standard, sub-eV to eV particle production
in the early Universe.
Amongst these are light sterile neutrinos (motivated by
popular interpretations of the LSND and the MiniBooNE results), and
hot dark matter axions (possible solution to the strong $CP$ problem).  The energy density residing in these additional 
light particles is conventionally parameterized as $\Delta N_{\rm eff}$, i.e., the effective number of additional 
``neutrino families''.  
 Phenomenologically, the presence of 
additional light particles has been shown in the past to exhibit considerable degeneracy with 
neutrino masses~\cite{Hannestad:2003xv} but could even lead to tighter constraints~\cite{Giusarma:2011ex}.  Using a combination of distance and clustering probes 
it is still possible to constrain $\sum m_\nu$ to the sub-eV level in this class of cosmological models~\cite{Hamann:2010bk}.
It should be noted that these scenarios generically predict modifications to the outcome of
big bang nucleosynthesis and thus can be independently constrained by observations of the
primordial light elemental abundances (e.g., \cite{Simha:2008zj}).

\item {\it Warm instead of cold dark matter:} \quad  Warm dark matter (WDM) scenarios invoke keV-mass particles 
in order to suppress the formation of dwarf galaxy-sized objects and potentially alleviate
the cusp problem in dark matter halos. The effects of replacing CDM with WDM are generally limited to the very small 
scales~\cite{Viel:2005qj}, and are not degenerate with light neutrino
masses.

\item {\it Inflation physics:} \quad  Popular extensions to the simplest description of the primordial 
perturbations include a running spectral index (i.e., a scale-dependent spectral index), the presence of a 
significant primordial gravitational wave background (a generic prediction of certain classes of inflation models), 
and isocurvature modes (from, e.g., multi-field inflation).  The latter two affect only the CMB
anisotropies at low multipoles and are not directly degenerate with neutrino masses.
A running spectral index can in principle mimic or offset to some extent the small-scale suppression in the matter power spectrum 
caused by free-streaming massive neutrinos.  However, running can be tightly constrained by the CMB
anisotropies.  Indeed, constraints on $\sum m_\nu$ from the WMAP 5-year data were already completely 
indepdendent of these additional parameters from inflation~\cite{Dunkley:2008ie}.

\item {\it Dynamical dark energy:} \quad  These scenarios involve replacing the cosmological constant with 
a fluid whose equation of state parameter $w$ satisfies  $w< -1/3$ and may additionally be time-dependent.  A popular 
realization, the so-called quintessence, uses a slowly rolling scalar field to achieve $w \simeq -1$.  
More elaborate variants usually involve some degree of coupling between the scalar field and other matter components.
The dark energy equation of state parameter $w$ was previously shown to exhibit considerable degeneracy 
with the neutrino mass~\cite{2003PhRvL..91d1301A,Hannestad:2005gj}.
However, in the post-BAO era, a combination of distance probes (e.g., BAO and Supernova Ia) can very effectively 
remove this degeneracy~\cite{Goobar:2006xz,Reid:2009nq}; use of CMB lensing information will also help in the future 
\cite{linder}. 
A more non-trivial issue concerns coupled dark energy scenarios and the possibility of their giving rise to 
scale-dependent clustering which may mimic or offset the effects of
neutrino masses.  This issue has yet to be explored in detail.

\item {\it Modified gravity:} \quad These scenarios, which modify general relativity at very 
large distances, are primarily constructed to explain the observed 
late-time accelerated expansion of the universe in lieu of a cosmological constant.  Phenomenologically
they share some similarities with the dynamical dark energy scenarios discussed above.
See \cite{linder1,linder2} for some discussion of 
the covariance of effects on the matter power spectrum from gravity modifications and from neutrino mass

\item  {\it Non-flat spatial geometry:} \quad  Flat spatial geometry is one of the pillars of the 
inflationary paradigm; relaxing the assumption of spatial flatness is perhaps the least theoretically
well-motivated extension to $\Lambda$CDM discussed here.  Phenomenologically, 
non-flat scenarios modify the distance-to-redshift relations in much the same way as the modifications encountered
in dynamical dark energy scenarios, and thus can be constrained using a combination of distance probes 
and Hubble parameter measurements~\cite{kom10}, leaving virtually no room for degeneracy with $\sum m_\nu$.

\end{enumerate}

Several classes of cosmological models that differ radically from $\Lambda$CDM have been proposed 
in the literature.  These include broken scale invariance~\cite{Blanchard:2003du} 
and void models that seek to explain current 
cosmological observations without invoking a phase of late-time accelerated expansion. 
While these models have found some success with certain subsets of the available data, it is generally difficult to
reconcile the simplest variants with all data sets.

From the above discussion, it is fair to conclude that neutrino mass limits from cosmology can be considered robust with respect to
reasonable modifications of the $\Lambda$CDM model.  Nonetheless, we stress again here that these limits
are necessarily derived from an inference process, and since the effects of neutrino mass on cosmological observables are 
purely gravitational, even a positive detection of hot dark matter in the future will not uniquely identify it as 
the neutrino with the correct quantum numbers.  However, should a future cosmological neutrino mass measurement find concordance 
with the outcome of experiments sensitive to mass differences and absolute masses, then it would provide an unambiguous confirmation of the $\Lambda$CDM paradigm.

\section{Conclusion}

It has often been said that cosmological probes of neutrino masses are complementary to terrestrial probes. On the simplest level, this is obvious in that cosmology is sensitive to the sum of the neutrino masses and terrestrial experiments probe either mass differences or different linear combinations of the masses. On a deeper level, though, information from terrestrial experiments will be a driving force in cosmology for years to come. 
Cosmology now has the almost unique opportunity/challenge to measure a fundamental physics parameter with a wide variety of probes. The challenge is whether each can fit its data with the plain vanilla $\Lambda$CDM model (which has been so successful to date) plus $m_\nu$. If these probes converge on the value of $m_\nu$ measured by terrestrial experiments, it will constitute one of the great triumphs of modern cosmology. If not, then we will have evidence for new physics: quintessence, modified gravity, non-standard inflation, curvature, or other possibilities whose richness is only hinted at by the parameters used to describe them. Sorting through the possibilities will likely require new probes and drive the field for generations.

\end{document}